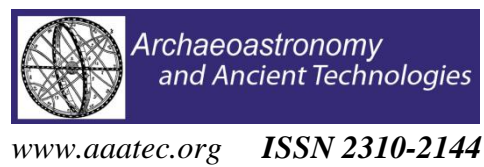

# Marks of Heliacal Rising of Sirius on the Sundial of the Bronze Age

**Larisa N. Vodolazhskaya[1], Anatoliy N. Usachuk[2], Mikhail Yu. Nevsky[3]**

[1] Southern Federal University (SFU), Rostov-on-Don, Russian Federation;
E-mails: larisavodol@aaatec.org, larisavodol@yahoo.com
[2] Donetsk Regional Museum, Donetsk, Ukraine; E-Mail: doold@mail.ru
[3]Southern Federal University (SFU), Rostov-on-Don, Russian Federation; E-mail: unevsky@sfedu.ru

**Abstract**

The article presents the results of interdisciplinary research made with the help of archaeological, physical and astronomical methods. The aim of the study were analysis and interpretation corolla marks of the vessel of the Late Bronze Age, belonging to Srubna culture and which was found near the Staropetrovsky village in the northeast of the Donetsk region. Performed calculations and measurements revealed that the marks on the corolla of Staropetrovsky vessel are marking of horizontal sundial with a sloping gnomon.

Several marks on the corolla of the vessel have star shape. Astronomical calculations show that their position on the corolla, as on "dial" of watch, indicates the time of qualitative change the visibility of Sirius in the day its heliacal rising and the next few days in the Late Bronze Age at the latitude of detection of Staropetrovsky vessel. Published in the article the results of astronomical calculations allow to state that astronomical year in the Srubna tradition began with a day of heliacal rising of Sirius.

**Introduction**

In 1985, near the village of Staropetrovsky (neighborhood Yenakiyevo, Donetsk region, Ukraine) a clay pot, owned Srubna culture and dating XV–XIV centuries BC, was found in a ruined barrow [1–3].

Its uniqueness lies in the combination groups of signs printed on the outer and inner surface of the vessel. Especially rare is the label on the inside of the vessel, which is a vertical row of nail marks and found themselves the markup of accumulative type water clock [4].

The cut of corolla of Staropetrovsky vessel decorated with nail imprints. Location nail impressions on the corolla is anisotropic, but has sufficiently clear axial symmetry about an axis passing through the centers of the sectors with the lowest and highest frequency of nail impressions (Fig. 1).

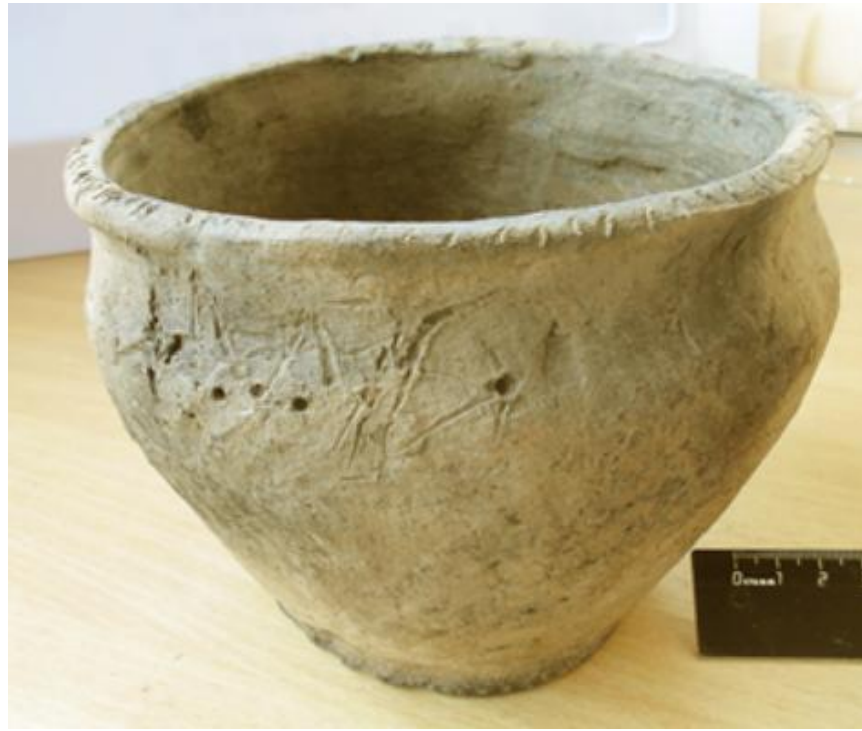

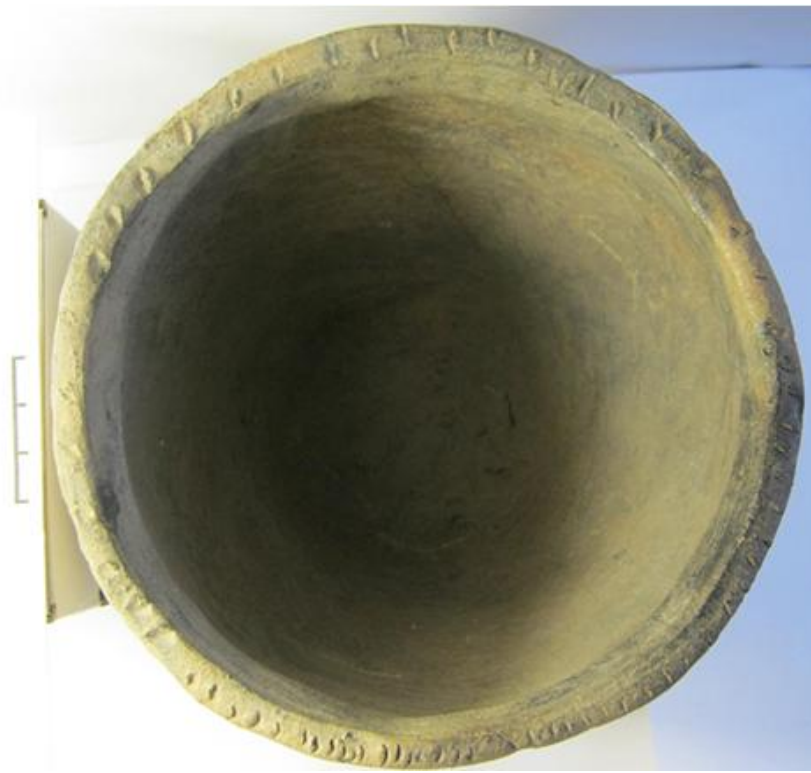

**Figure 1.** Vil. Staropetrovsky, ruined barrow, photo of vessel with the marks on the inner surface: ***a*** – side view; ***b*** – top view (Photo by A.N. Usachuk, 2014).

Location wells (nail impressions) on the corolla of the vessel resembles location hour markers of sundial. To prove that the wells of corolla can really be a sundial hour markers, we conducted interdisciplinary research by means of complex natural science methods that have already proven to be effective in similar studies [5–12].

Taking into account that the vessel has a flat bottom, and in its operating position the plane of of the corolla will be close to horizontal, we decided to compare the wells of the corolla with hour marks of horizontal sundial[1]. Comparison with marks of analemmatic sundial [13, 14] (horizontal sundial with vertical gnomon) have not yielded positive results. Then we attempted to compare them with the hour markers of horizontal sundial with sloping gnomon, which led to very interesting results.

If we shift attachment point of the gnomon away from the center of the circular "dial" that location of hour marks will have a higher density on the side of "dial" the nearest to that place. Since operating range of sundial corresponds daytime, it is logical to assume that the wells are located with a minimum density, allowing to determine the time as accurately as possible (maximum resolution), corresponds to the northern side of the "dial". In this case, the wells are located at maximum density, corresponds to the opposite – south side of "dial". Location of "southern" wells corresponds to each hour only formally, because they are located in non–working zone of a sundial, where the shadow of the gnomon can't get there. Accordingly, for "southern" wells do not need the same high resolution as for the "northern" wells, that allows to reduce the distance between them.

Coordinate plane with hour lines of horizontal sundial with a sloping gnomon, offset toward the South, is shown in Figure 2.

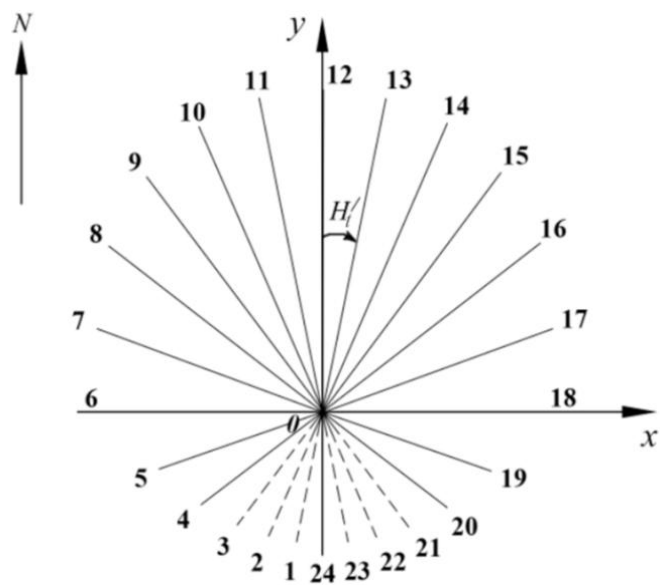

**Figure 2.** The coordinate plane with hour lines of horizontal sundial with a sloping gnomon: $H'_t$ – the angle between the meridian line and the hour line. $H'_{12}=0°$, $H'_6=-90°$, $H'_{18}=90°$. The dotted line marked hour lines not working range of a sundial from 21 o'clock to 3 o'clock, according to local true solar time. *N* – True North.

---

[1] At the same time we assumed that such markings originally certainly should be placed on the template, for example, on a wooden flat surface, and only then copied to the corolla, during the manufacture of the vessel.

When we determine the approximate orientation of the vessel and the axis of symmetry, on which should be placed a gnomon, we tried to more accurately determine the location of its attachment. To this end we conducted a computer experiment using graphic editor Adobe Photoshop CS5. For the experiment we have created a drawing of the corolla using photos of the vessel. Standing on a horizontal plane vessel was photographed from above, with a height of about one meter (of four times the diameter of the corolla, approximately), with the camera lens directed vertically downwards and arranged above the geometric center of the corolla vessel. This will minimize distortions arising when photographing.

In the experiment, a graphic layer with applied calculated hour lines was compared with the layer containing the photo and drawing marks of corolla of the vessel. To search for a possible site of attachment of the inclined gnomon we applied the method of morphological analysis of Zwicky, based on the selection of all possible solutions for individual parts of the task (the so–called morphological features that characterize the device) and then systematically getting their combinations [15]. As a device we considered staropetrovsky vessel corolla, as horizontal sundial ("dial") with a sloping gnomon. As the separate parts were considered: complex marks of corolla as hour markers and inclined gnomon (place attachment and orientation). In accordance with the method of morphological analysis, we carried out an exhaustive search of all possible variants of the provisions of the site of attachment of the gnomon and its orientation. The possible mechanism of fixing a gnomon at this stage of solving the problem we have not considered. In accordance with the method of morphological analysis, we conducted an exhaustive search of all the options of possible mounting points of gnomon near the symmetry axis of the corolla marks. Layer with hour lines moved us in a graphics editor vertically and horizontally (height and width of the drawing), in steps of 1 mm, and rotated relative to the center, hour lines in increments of 1º.

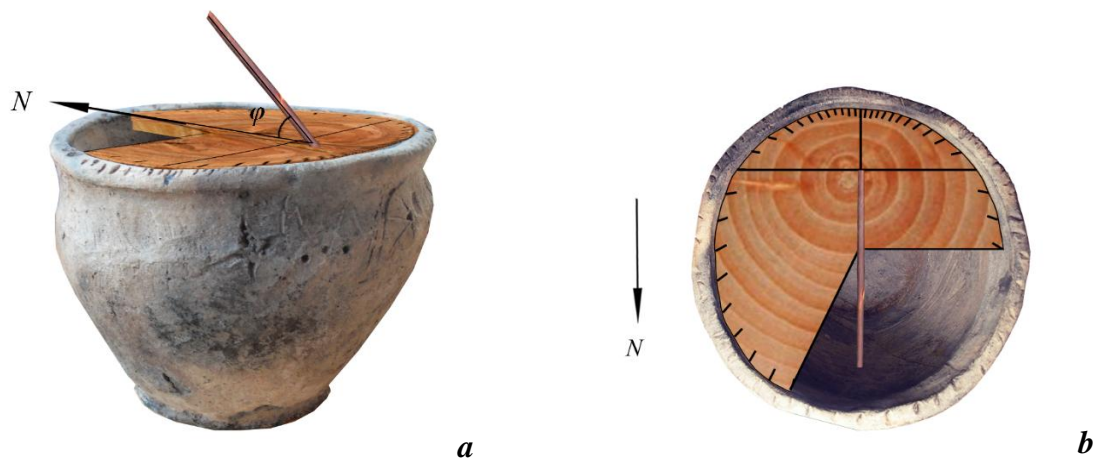

*a*                                                                                          *b*

**Figure 3.** Vil. Staropetrovsky, ruined barrow, reconstruction of the horizontal sundial, with a sloping, gnomon (made on the basis of photographs of the vessel). In the upper part of the vessel – wooden disc with applied accurate hour markers, fortified gnomon and the opening for receiving water into the vessel. $N$ – direction to the True North, $\varphi$ – latitude of Staropetrovsky village.

Revealed by us the most likely place attachment is characterized by the ratio of the distance from the gnomon to the nearest edge of the corolla to the distance from the gnomon to the maximum the distant edge of the corolla how to 1: 4. In this case, part of the corolla with often located small marks was oriented to the South and with sparse big marks – to the North. We believe that the most likely mount of the gnomon was associated with its installation on discoid wooden cover, which was placed in the upper part of the vessel. In our view, the cover could further have an independent markup corresponding hour lines, allows the correct combine the cover with the wells on the corolla and with a hole for receiving water for the simultaneous operation of the vessel and as sundial and as a water clock (Figure 3).

Calculation of hour angles for horizontal sundial, with a sloping, gnomon held us by the formulas [16, 17]:

$$H_t^{'} = arctg(\sin\varphi \cdot tg(H_t))$$

or

$$H_t^{'} = arctg(\sin\varphi \cdot tg(15 \cdot (t-12))), \text{ при } t \in [6; 18] \quad (1)$$

$$H_t^{'} = arctg(\sin\varphi \cdot tg(15 \cdot (t-12))) - 180, \text{ при } t \in [0; 6[ \quad (2)$$

$$H_t^{'} = arctg(\sin\varphi \cdot tg(15 \cdot (t-12))) + 180, \text{ при } t \in ]18; 24] \quad (3)$$

where $H_t = 15^0 \cdot (t-12)$ – the hour angle of the Sun (for half day $H_{12} = 0^0$), $t$ – time, $H'_t$ – the angle between the meridian line and the hour line of sundial (fig. 19), $\varphi$ – latitude. The results of our calculations by formulas 14–16 for latitude of Staropetrovsky village *Lat* = 48° 13' N are shown in Table 1.

**Table 1.** Calculated hour lines of horizontal sundial (from 6 am to noon). $H_t$ – hour angle of the Sun, $H'_t$ – calculated angle between the meridian line and hour line, $t$ – time.

| $t$, (hour) | 0 | 0.5 | 1 | 1.5 | 2 | 2.5 | 3 | 3.5 | 4 | 4.5 | 5 | 5.5 |
|---|---|---|---|---|---|---|---|---|---|---|---|---|
| $H_t$, (⁰) | -180.0 | -172.5 | -165.0 | -157.5 | -150.0 | -142.5 | -135.0 | -127.5 | -120.0 | -112.5 | -105.0 | -97.5 |
| $H'_t$, (⁰) | -180.0 | -174.4 | -168.7 | -162.8 | -156.7 | -150.2 | -143.3 | -135.8 | -127.7 | -119.1 | -109.8 | -100.0 |
| $t$, (hour) | 6.0 | 6.5 | 7.0 | 7.5 | 8.0 | 8.5 | 9.0 | 9.5 | 10.0 | 10.5 | 11.0 | 11.5 |
| $H_t$, (⁰) | -90.0 | -82.5 | -75.0 | -67.5 | -60.0 | -52.5 | -45.0 | -37.5 | -30.0 | -22.5 | -15.0 | -7.5 |
| $H'_t$, (⁰) | -90.0 | -80.0 | -70.2 | -60.9 | -52.3 | -44.2 | -36.7 | -29.8 | -23.3 | -17.2 | -11.3 | -5.6 |
| $t$, (hour) | 12.0 | 12.5 | 13.0 | 13.5 | 14.0 | 14.5 | 15.0 | 15.5 | 16.0 | 16.5 | 17.0 | 17.5 |
| $H_t$, (⁰) | 0.0 | 7.5 | 15.0 | 22.5 | 30.0 | 37.5 | 45.0 | 52.5 | 60.0 | 67.5 | 75.0 | 82.5 |
| $H'_t$, (⁰) | 0.0 | 5.6 | 11.3 | 17.2 | 23.3 | 29.8 | 36.7 | 44.2 | 52.3 | 60.9 | 70.2 | 80.0 |
| $t$, (hour) | 18.0 | 18.5 | 19.0 | 19.5 | 20.0 | 20.5 | 21.0 | 21.5 | 22.0 | 22.5 | 23.0 | 23.5 |
| $H_t$, (⁰) | 90.0 | 97.5 | 105.0 | 112.5 | 120.0 | 127.5 | 135.0 | 142.5 | 150.0 | 157.5 | 165.0 | 172.5 |
| $H'_t$, (⁰) | 90.0 | 100.0 | 109.8 | 119.1 | 127.7 | 135.8 | 143.3 | 150.2 | 156.7 | 162.8 | 168.7 | 174.4 |

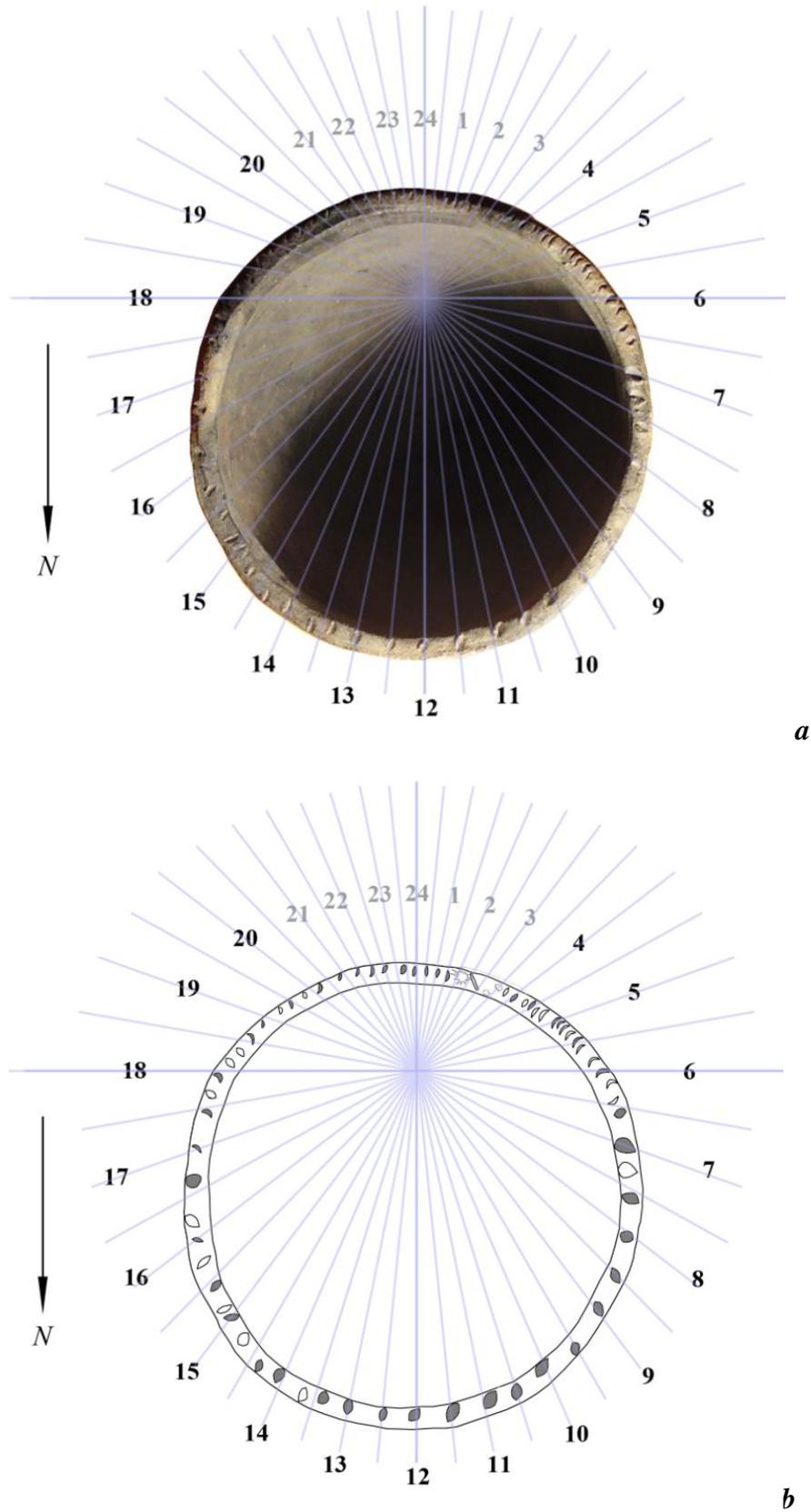

**Figure 4.** Vil. Staropetrovsky, ruined barrow, vessel with the marks on the inner surface: *a* – photo of the corolla with marks; *b* – drawing of the corolla with marks. Blue lines indicate hour lines of horizontal sundial with a sloping gnomon. The numbers plotted clocks corresponding hour lines. Gray color indicates the marks, located in immediate vicinity of the hour lines. *N* – True North.

Optimal results combining of calculated hour angles with the marks on corolla of Staropetrovsky vessel is shown in Figure 4. Number of marks on the rim of the vessel exceeds the number of calculated hour lines, however, clearly visible coincidence hour lines with wells in the range from eight to 13 o'clock.

To test our hypothesis on the implementation of marks of the corolla of Staropetrovsky vessel functions hour markers sundial, all marks of the corolla were numbered and for each of them (approximately for their geometric centers), we measured the angular distance relative to meridian line (Tab. 2).

**Table 2.** The angular distance from the marks of the corolla relatively meridian line. $Z$ – number of mark; $H'_z$ – angular distance relative to the meridian line.

| Z | $H'_z$, (°) | Z | $H'_z$, (°) | Z | $H'_z$, (°) |
|---|---|---|---|---|---|
| 1 | -180.0 | 23 | -86.6 | 46 | 38.7 |
| 2 | -174.7 | 24 | -81.1 | 47 | 42.9 |
| 3 | -168.3 | 25 | -78.2 | 48 | 47.9 |
| 4 | -162.2 | 26 | -69.7 | 49 | 52.2 |
| 5 | -153.7 | 27 | -64.7 | 50 | 56.1 |
| 6/1 | -149.0 | 28 | -59.0 | 51 | 63.6 |
| 6/2 | -143.1 | 29 | -51.6 | 52 | 70.5 |
| 7 | -139.5 | 30 | -44.0 | 53 | 78.7 |
| 8 | -136.2 | 31 | -37.2 | 54 | 83.5 |
| 9 | -132.7 | 32 | -29.5 | 55 | 88.2 |
| 10 | -127.4 | 33 | -22.8 | 56 | 92.5 |
| 11 | -122.4 | 34 | -17.2 | 57 | 96.2 |
| 12 | -119.9 | 35 | -12.5 | 58 | 101.2 |
| 13 | -117.0 | 36 | -6.0 | 59 | 106.8 |
| 14 | -114.3 | 37 | 0.3 | 60 | 114.0 |
| 15 | -109.9 | 38 | 5.6 | 61 | 118.1 |
| 16 | -108.5 | 39 | 11.5 | 62 | 124.2 |
| 17 | -106.3 | 40 | 15.8 | 63 | 130.7 |
| 18 | -104.0 | 41 | 19.3 | 64 | 140.7 |
| 19 | -100.8 | 42 | 24.2 | 65 | 149.1 |
| 20 | -98.1 | 43 | 28.0 | 66 | 156.0 |
| 21 | -93.6 | 44 | 32.7 | 67 | 162.7 |
| 22 | -89.8 | 45 | 36.7 | 68 | 172.4 |

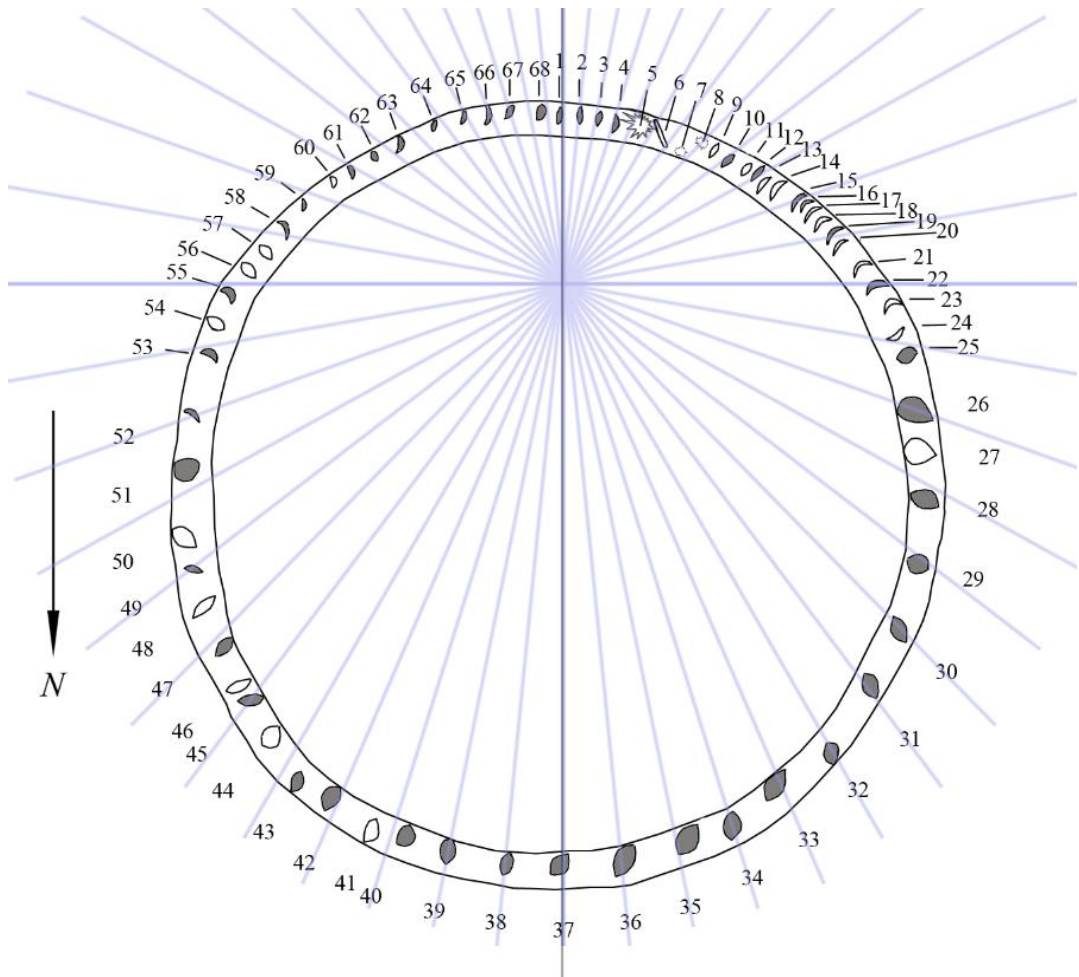

**Figure 5.** Vil. Staropetrovsky, ruined barrow, vessel with the marks on the inner surface, drawing of the corolla with marks. Blue lines indicate hour lines of horizontal sundial with a sloping gnomon. Digits – the numbers of marks. Gray color indicates the mark located in immediate vicinity from the hour lines. *N* – True North.

For all marks, through which pass or nearby from which there are hour lines (Fig. 5), we calculated the absolute errors of measured marks time relative of calculated hour lines by the formula 4:

$$\Delta t'_z = \frac{\Delta H'_z}{\Delta H'_t} \cdot 30 \qquad (4)$$

where $\Delta t'_z$ – absolute error of the measured time–stamped relative to calculated hour lines, $\Delta H'_t = H'_t - H'_{t+0.5}$ – the angular distance between adjacent hour lines closest to the mark with the number Z (see. tab. 11 and 12), $\Delta H'_z = H'_t - H'_z$ – the difference between the nearest to the mark Z hour line $H'_t$ and angular distance of mark $H'_z$ relatively meridian line.

The results of calculations according to formula 4 are shown in table 3. Contemplated marks allocated in the figure 5 gray color. Marks of corolla, which are not explicitly refer to the marks of sundial, shown in Figure 5 without the color fill. It is possible that these marks, except marks 5–9 (see below), may be used to masking technology of measurements of time and prevented the use of a vessel as a sundial without participation of its owner.

**Table 3.** The absolute errors measured by the marks of time relative to the calculated hour lines. $Z$ – number of the mark, $H'_z$ – angular distance mark relative meridian line, $T$ – time, $H'_t$ – hour line closest to the mark $Z$, $\Delta H'_z$ – difference between the hour line $H'_t$ and angular distance of mark $H'_z$, $\Delta t'_z$ – absolute error of the measured time–stamped, nearest relative to mark, the time line.

| Z | $H'_z$, (°) | T, (hour) | $H'_t$, (°) | $\Delta H'_z$, (°) | $\Delta t'_z$, (min) |
|---|---|---|---|---|---|
| 1 | -180.0 | 0.0 | -180.0 | 0.0 | 0.0 |
| 2 | -174.7 | 0.5 | -174.4 | 0.3 | -1.6 |
| 3 | -168.3 | 1.0 | -168.7 | -0.4 | 2.1 |
| 4 | -162.2 | 1.5 | -162.8 | -0.6 | 3.1 |
| 10 | -127.4 | 4.0 | -127.7 | -0.3 | 1.1 |
| 12 | -119.9 | 4.5 | -119.1 | 0.8 | -2.8 |
| 15 | -109.9 | 5.0 | -109.8 | 0.1 | -0.3 |
| 19 | -100.8 | 5.5 | -100.0 | 0.8 | -2.5 |
| 22 | -89.8 | 6.0 | -90.0 | -0.2 | 0.6 |
| 25 | -78.2 | 6.5 | -80.0 | -1.8 | 5.4 |
| 26 | -69.7 | 7.0 | -70.2 | -0.5 | 1.5 |
| 28 | -59.0 | 7.5 | -60.9 | -1.9 | 6.1 |
| 29 | -51.6 | 8.0 | -52.3 | -0.7 | -2.4 |
| 30 | -44.0 | 8.5 | -44.2 | -0.2 | -0.7 |
| 31 | -37.2 | 9.0 | -36.7 | 0.5 | -2.0 |
| 32 | -29.5 | 9.5 | -29.8 | -0.3 | -1.3 |
| 33 | -22.8 | 10.0 | -23.3 | -0.5 | -2.3 |
| 34 | -17.2 | 10.5 | -17.2 | 0.0 | 0.0 |
| 35 | -12.5 | 11.0 | -11.3 | 1.2 | 6.1 |
| 36 | -6.0 | 11.5 | -5.6 | 0.4 | 2.1 |
| 37 | 0.3 | 12.0 | 0.0 | -0.3 | 1.6 |
| 38 | 5.6 | 12.5 | 5.6 | 0.0 | 0.0 |
| 39 | 11.5 | 13.0 | 11.3 | -0.2 | 1.1 |
| 40 | 15.8 | 13.5 | 17.2 | 1.4 | -7.2 |
| 42 | 24.2 | 14.0 | 23.3 | -0.9 | 4.4 |
| 43 | 28.0 | 14.5 | 29.8 | 1.8 | -8.3 |
| 45 | 36.7 | 15.0 | 36.7 | 0.0 | 0.0 |
| 47 | 42.9 | 15.5 | 44.2 | 1.3 | -5.2 |
| 49 | 52.2 | 16.0 | 52.3 | 0.1 | -0.4 |
| 51 | 63.6 | 16.5 | 60.9 | -2.7 | 9.3 |
| 52 | 70.5 | 17.0 | 70.2 | -0.3 | 1.0 |
| 53 | 78.7 | 17.5 | 80.0 | 1.3 | -4.0 |
| 55 | 88.2 | 18.0 | 90.0 | 1.8 | -5.4 |
| 58 | 101.2 | 18.5 | 100.0 | -1.2 | 3.6 |
| 59 | 106.8 | 19.0 | 109.8 | 3.0 | -9.2 |
| 61 | 118.1 | 19.5 | 119.1 | 1.0 | -3.2 |
| 62 | 124.2 | 20.0 | 127.7 | 3.5 | -12.1 |

| 63 | 130.7 | 20.5 | 135.8 | 5.1 | -19.0 |
|----|-------|------|-------|-----|-------|
| 64 | 140.7 | 21.0 | 143.3 | 2.6 | -10.4 |
| 65 | 149.1 | 21.5 | 150.2 | 1.1 | -4.8  |
| 66 | 156.0 | 22.0 | 156.7 | 0.7 | -3.2  |
| 67 | 162.7 | 22.5 | 162.8 | 0.1 | -0.5  |
| 68 | 172.4 | 23.5 | 174.4 | 2.0 | -10.5 |

On the "dial" of sundial, marked for the range of 24 hours every half hour, must be 48 hour marks. If we assume that each hour mark corresponds to one degree of circle, then probability of a random match the well – marks (center of well) on the corolla with hour marks can be calculated by the formula 5:

$$p=m/n \tag{5}$$

where p – the probability of event A in a statistical sense, m – the average number of occurrences of the event A, n – the number of tests.

Let the random selection degree, corresponding to the hour mark, on the circumference when applied marks on the corolla – round "dial" of a sundial – will be treated as an event *A*. Then *m=48* – the total number of degrees of the circumference, corresponding to all hour markers, and *n=360* – the total number of degrees of a circumference. Calculated by the formula 5 probability of random coincidence mark (center of mark) on the corolla with the hour mark is *p=m/n=48/360≈0.13=13%*.

In case of accidental putting to the corolla of Staropetrovsky vessel 68 wells (*n = 68*), the number of wells, which coincided with the 48 hour markers, must be equal to *m=68·0.13≈9* wells. However, on the corolla of Staropetrovsky vessel with hour marks coincides not nine, but 33 wells with an accuracy of ≈±1° (relative of the wells centers), which is almost four times higher than the probability of random coincidences.

Also, if it is assumed a random marking of corolla, then when selecting 11 wells of corolla (*n=11*) consecutive, comply with hour marks should only be *m=11·0.13=1.5≈2* wells, and in the range from 8 to 13 o'clock, all *11* wells correspond hour marks with an accuracy of ≈±1°, which is more than five times higher than the probability of random coincidences.

Thus, the number of wells on the corolla of the Staropetrovsky vessel coinciding with the hour markers, indicative of the fact that the marking of corolla was made not random way and with a high probability corresponds to marking of the horizontal sundial with sloping gnomon for the latitude of detection Staropetrovsky vessel.

The average absolute error of time measurement in the range of 8 to 13 o'clock $\Delta t'_{z\_mean}$ ≈1.8 minutes was minimal compared to other ranges, as well as the corresponding standard deviation σ≈1.6 minute, calculated according to a formula similar to the formula 4. I.e. in this range the time measurement could be made with maximum precision, that testifies to the key role this range in the technology of measuring time using the are described sundial. Mark 12 o'clock is in this range. With 12 o'clock due shortest shadow during daylight hours. Therefore, this range could be used for control the spatial orientation of the vessel, for control the rate of filling of the vessel and for measure deviations of the sundial from a water clock.

Analogously calculated by us the mean absolute error for the range, including all marks, $\Delta t'_{z\_mean}$≈3.9 minutes, standard deviation σ≈3.9 minutes.

The calculated average absolute error for the non−working time range of sundial from 20 to 4 o'clock has a maximum value in comparison with all other ranges: $\Delta t'_{z\_mean}$≈5.5 minutes and standard deviation σ≈5.7 minutes. This is an expected result for a range of marks that could not participate in the direct measurement of time by a sundial.

To working range of sundial, determined by the time of sunrise and sunset at the summer solstice, relate only marks from 4 to 20 o'clock. Calculated by us the average absolute error for this range: $\Delta t'_{z\_mean}$≈3.4 minutes, standard deviation σ≈3.1 minutes. The average absolute error for the time range, the limited time of sunrise and sunset at the equinoxes from 6 to 18 o'clock: $\Delta t'_{z\_mean}$≈3.1 minutes, standard deviation σ≈2.7 minutes. The average absolute error for the time range, the limited time of sunrise and sunset in the days of the winter solstice from 8 to 16 o'clock: $\Delta t'_{z\_mean}$≈2.7 minutes. The calculated standard deviation σ≈2.5 minutes. Given the results obtained, we can conclude that in the working range for the winter solstice, for equinox and for the summer solstice, accuracy of possible measurements of time with marks of the corolla was approximately identical and equal ≈3 minutes.

Thus, we believe that marks of the corolla of staropetrovsky vessel could serve as hour markers of sundial with a sloping gnomon. With its help in the time range from 4 to 20 o'clock may be conducted measurement of time with an average accuracy of 3.4 minutes. This accuracy is close accuracy of measurement of time by Egyptian vertical sundial with sloping gnomon the same epoch ≈3.6 min [12]. This fact is indirect evidence about the similarities and mediated transfer of technology of measurement of time between these regions.

**Marks of heliacal rising of Sirius**

On the corolla of Staropetrovsky vessel almost all marks are putting using nail impressions. However, marks with numbers 5–8 are from different (Fig. 5). Mark number 5 resembles a star and made using scribe lines from a central point by a thin sharp object, apparently pointed reed (Fig. 6). Similar acute sticks – typical tools of pottery technology of **Srubna** culture [19].

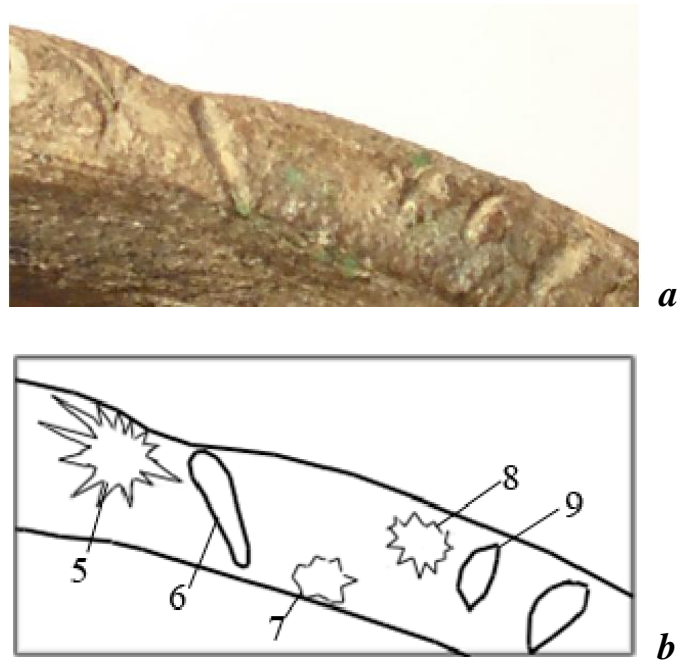

**Figure 6.** Vil. Staropetrovsky, ruined barrow, vessel with the marks on the inner surface, fragment of corolla with marks 5–9: ***a*** – photo, ***b*** – drawing. Digits – the numbers of marks.

Starlike mark 5 is similar to the image of the stars. Similar images of stars are found, for example, on Mesopotamian cuneiform tablets and seals (Fig. 7).

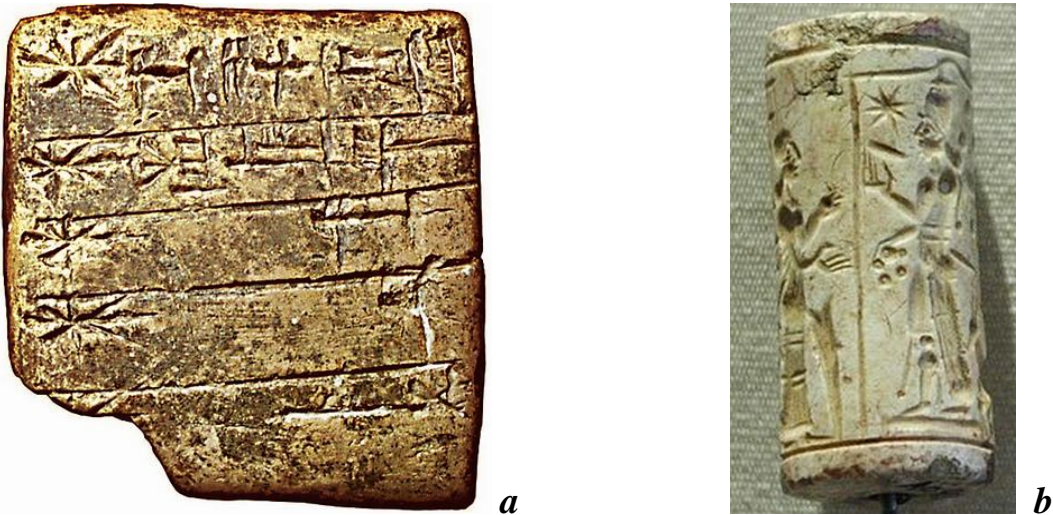

*a*    *b*

**Figure 7.** Examples of images of stars: *a* – Sumerian clay cuneiform tablet with a list of the gods (about 2400 BC)[2], whose names begin with the sign DINGIR (star, sky, time, deity), developed from pictographic image of a star, *b* – Mesopotamian limestone cylinder seal with an image of service to the Sun–God Shamash (about 1000 BC)[3].

A mark with the number 6 is forward slash in the form of imprint wand is likely on the reed. Marks with numbers 7 and 8 resemble on stars, but have much smaller dimensions than mark 6 (fig. 6). These marks are made by means of pressing a thin sharp object. By a group starlike marks on the corolla of Staropetrovsky vessel on its side surface the star image, followed by a complicated composition, is located also (Fig. 8).

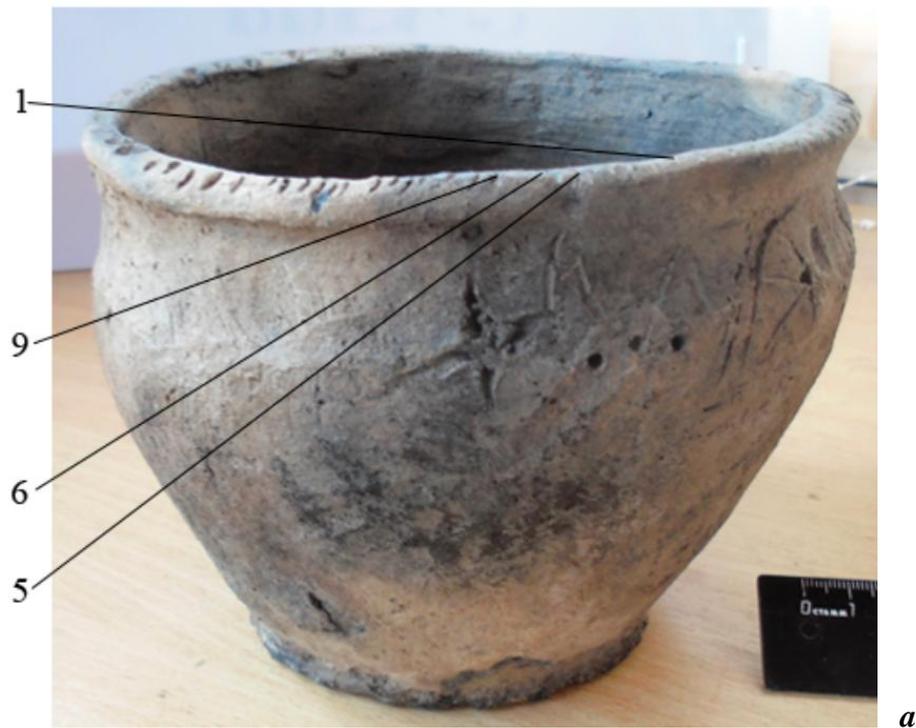

*a*

---

[2] https://ru.wikipedia.org/wiki/Dingir#/media/File:Sumerian_MS2272_2400BC.jpg
[3] http://cartelen.louvre.fr/cartelen/visite?srv=car_not&idNotice=26739

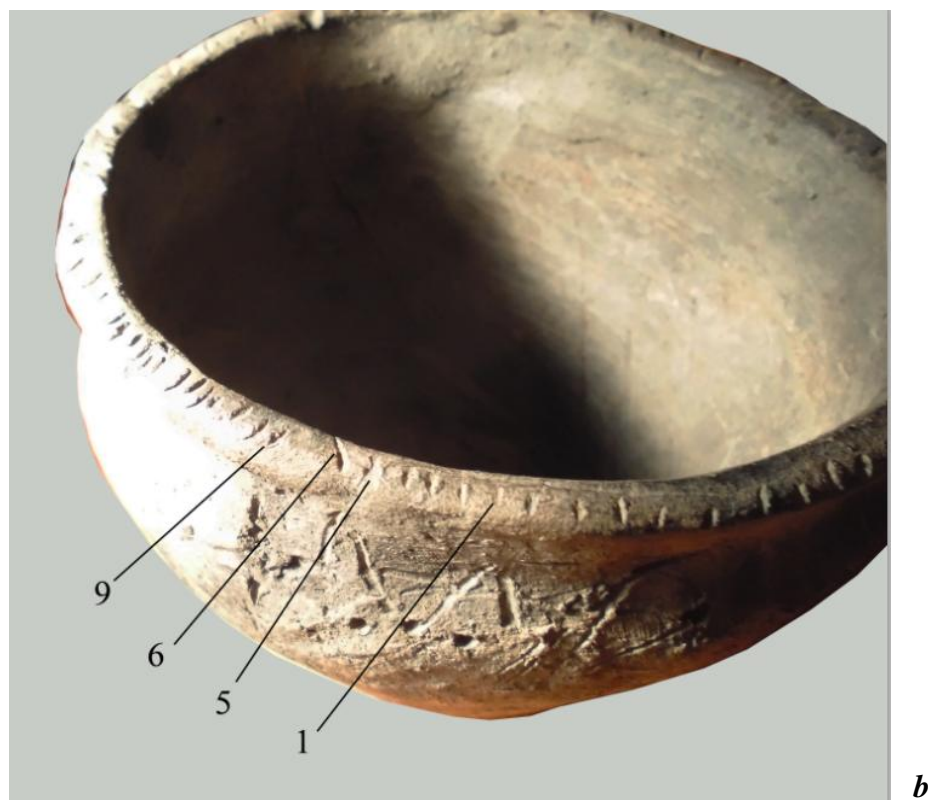

***b***

**Figure 8.** Vil. Staropetrovsky, ruined barrow, vessel with the marks on the inner surface, a fragment of corolla with marks 5–9: ***a*** − side view, ***b*** − the top view. Digits − the numbers of marks.

We assumed that the image of a star on the corolla of the vessel and on its side surface could be associated with one and the same star. It is known that in the Babylonian astrolabes (1200 BC) Heliacal rising of various stars corresponded to the calendar months [20].

The brightest star in the night sky in the northern hemisphere is Sirius. It is visible on almost any latitude up to 73º *N*, and it is the brightest star in the entire night sky[4], with magnitude −1.46$^m$ [21, p. 510]. In Mesopotamia, at least from the VII century BC, regularly recorded heliacal risings, heliacal settings and akronical risings (the latter rising stars, visible after sunset) Sirius [22, 23], and in ancient Egypt from III millennium BC namely the heliacal rising of Sirius was particularly important. In ancient Egypt, from III millennium BC from heliacal rising[5] of Sirius began to count the beginning of the year, which initially coincide with the summer solstice and the flooding of the Nile [21, p. 41].

The phenomenon of heliacal rising of Sirius is characterized by a height[6] of Sirius from about +2º to +3º, and the height of the Sun from -8º to -6º [24]. Ancient Egyptian calendar, in which the beginning of the year corresponded date of heliacal rising of Sirius, was first introduced in 2767 BC in Memphis (29°50′40″ N, 31°15′03″ E) [25, 26]. In this year the heliacal rising of Sirius coincided with the summer solstice. In astronomy program RedShift−7 Advanced date for the summer solstice this year falls on 16.07 (for the modern age, this date corresponds to June 21, 2015 relative to the day of the summer solstice). Using the same program we determined that the height of Sirius reached +3º, with simultaneous height Sun ≈-6º on this day. The height of the Sun was determined relative to the center of the visible solar disk.

---

[4] https://en.wikipedia.org/wiki/List_of_brightest_stars
[5] The first appearance of the star before sunrise in a year
[6] The height *h* is called an arc from the plane of the mathematical horizon before the direction to the object.

Given that the parameters of the position of the Sun and Sirius, on the day of heliacal rising of the latter, are due the objective features of visibility of Sirius, we have used these parameters and to determine the days of heliacal settings and akronical risings (the latter rising stars, visible after sunset) Sirius. For these days, by analogy with the heliacal rising we took Sirius height ≈ +3º, at the height of the sun ≈-6º. Date and time of heliacal settings and akronical risings were calculated by us for the 1400 BC using astronomical software RedShift 7 Advanced.

At the latitude of the Staropetrovsky village for height Sirius h≈ +3º and the height of the Sun h≈-6º heliacal rising of Sirius observed 9.08.1400 BC (for the modern age corresponds to July 25, 2015 regarding the date of the summer solstice[7]), approximately at 3:40 (local solar time with respect to noon), heliacal setting – 8.11.1400 BC at 6:23 and akronical rising of Sirius – 17.01.1400 BC at 16:51. Mark of the star on the corolla of the vessel is adjudged in the time range from about two to four o'clock in the morning, which corresponds to the heliacal rising of Sirius.

Heliacal rising of Sirius in 1400 BC observed 9 August in a range approximately, ±100 years. Calculated by means program RedShift 7 Advanced time $T_s$ of rising of Sirius for height of Sirius $h_s$ equal to 0º, +2º, +3º above the horizon, and after analyzing it, we found that it is close to time $T_z$, indicated by the marks 7, 8 and 9. Calculation $T_z$ – time, corresponding to the mark with a given number Z, made by the formulas:

$$\Delta t'_s = \frac{\Delta H'_s}{\Delta H'_t} \cdot 30 \qquad (6)$$

$$T_Z = T_s + \Delta t'_s \qquad (7)$$

where $\Delta t'_s$ – difference time, corresponding to mark with the number Z, from the time of astronomical event, associated with the apparent position of Sirius, $\Delta H'_t = H'_t - H'_{t+0.5}$ – the angular distance between adjacent hour lines, nearest on time to astronomical events (see. tab. 1), $\Delta H'_s = H'_s - H'_z$ – the difference between the hour line $H'_s$, corresponding to astronomical event, and the angular distance $H'_z$, on which located a mark with the number Z relative to the noon line, $T_z$ – time, corresponding to the mark with the number Z, $T_s$ – time of astronomical event associated with Sirius.

The results of our calculations according to formulas 6 and 7 shown in Table 4.

**Table 4.** Time of rising of Sirius near the date of the heliacal rising and the time corresponding to the marks 5–9 on the corolla of staropetrovsky vessel. Z – number of the mak, $h_s$ – height of Sirius, $h_{sun}$ – height of the Sun, $T_s$ – time of astronomical event associated with the apparent position of Sirius, $H_s'$ – the angular distance, on which should be mark relative to the midday line, for the time of a specific astronomical event, $H'_z$ – angular distance, at which locate the mark with number Z respect of meridian line, $\Delta H'_s$ – the difference between the angular distance $H_s'$ and angular distance $H'_z$, $T_z$ – time corresponding to the mark with the number Z, $\Delta t'_s$ – absolute deviation of time corresponding to the mark with the number Z, from the time of an astronomical event that is associated with the apparent position of Sirius.

---

[7] At the latitude of the village Staropetrovsky heliacal rising of Sirius in 2015 will be observed on August 21.

| Z | $h_s$, (º) | $h_{sun}$, (º) | Date (BC) | $T_s$, | $H_s'$, (º) | $H_z'$, (º) | $\Delta H_s'$, (º) | $\Delta t'_s$, (min) | $T_z$, |
|---|---|---|---|---|---|---|---|---|---|
| 5 | +0º04′ | -20º03′ Astronomical night | 26.08.1400 | 2:13 | -153.9 | -153.7 | -0.2 | 0.9 | 2:13.9 |
| 6/1 | +3º07′ | 17º58′ The beginning of astronomical twilight[8] | 26.08.1400 | 2:34 | -149.3 | -149.0 | -0.3 | 1.4 | 2:35.4 |
| 6/2 | +3º04′ | -12º26′ The beginning of nautical twilight | 18.08.1400 | 3:05 | -142.1 | -143 | 0.9 | -3.6 | 3:01.4 |
| 7 | +0º00′ | -8º36′ | 9.08.1400 | 3:20 | -138.4 | -139.5 | 1.1 | -4.4 | 03:15.6 |
| 8 | +2º00′ | -6º59′ | 9.08.1400 | 3:33 | -135 | -136.2 | 1.2 | -4.5 | 03:28.5 |
| 9 | +3º00′ | -6º06′ The beginning of civil twilight | 9.08.1400 | 3:40 | -133.2 | -132.7 | -0.5 | 1.9 | 3:41.9 |
| 9 | +2º55′ | -20º03′ Astronomical night | 8.08.0900 | 3:45 | -131.9 | -132.7 | 0.8 | -3.0 | 3:41.9 |
| 9 | +3º01′ | 17º58′ The beginning of astronomical twilight | 10.08.1900 | 3:37 | -134.0 | -132.7 | -1.3 | 4.8 | 3:41.9 |
| 9 | +2º57′ | -12º26′ The beginning of nautical twilight | 11.08.2400 | 3:34 | -134.8 | -132.7 | -2.1 | 7.8 | 3:41.9 |

Analyzing the results of the calculations are presented in Table 14, we can conclude that the mark of 5–9, indeed, noted time points that correspond to characteristic visible the provisions of Sirius in a sky near the date of its heliacal rising on the corolla of the vessel as on the dial. Traced by the time of rising of Sirius using the sundial was impossible. It was possible only by using a water clock.

The mark 7 is the earliest by date and time from 5–9 marks. It is a small star–shaped mark, which is located at the inner edges of the corolla. It corresponds to the time, when Sirius must be on the line of horizon. The Sun located approximately by 9º below the horizon at the same time. This position of the Sun corresponds to nautical twilight, which are characterized by good visibility only the brightest – navigational stars, which include Sirius. Most stars are almost impossible to see in close proximity to the horizon, but Sirius is the brightest star in the night sky, what makes the possibility of such observations more probable under favorable weather conditions. In addition, the neighborhood of the Staropetrovsky village characterized by a flat terrain, one not interfering the observations near the horizon (Fig. 9). Perhaps the heliacal rising of Sirius was observed in the direction of a small (1º– 2º) natural lowering of the horizon relative to the main line of the horizon as in the photo (Fig. 9), which makes possible the observation at time corresponding mark 7.

---

[8] At the latitude of the village Staropetrovsky such astronomical situation will observe in 2015, 4 September.

Mark 8 is the second star–shaped small mark and located at the outer edge of the corolla. It corresponds to the height of Sirius +2º. At the same time, the Sun is approximately at 7º below the horizon, which also corresponds to the nautical twilight. The fact that the mark 8 is similar in shape and size to the mark 7, but is farther away, as it were, above the site of attachment of the gnomon, can be seen as confirmation that the two marks represent one and the same object in the sky – the star, but at different heights above the horizon. The small size of these marks, in this case, can be interpreted as a weak visibility of stars associated with its proximity to the horizon and, at the same time, with the proximity of its observations on the time to completion of nautical twilight, when the total coverage increases.

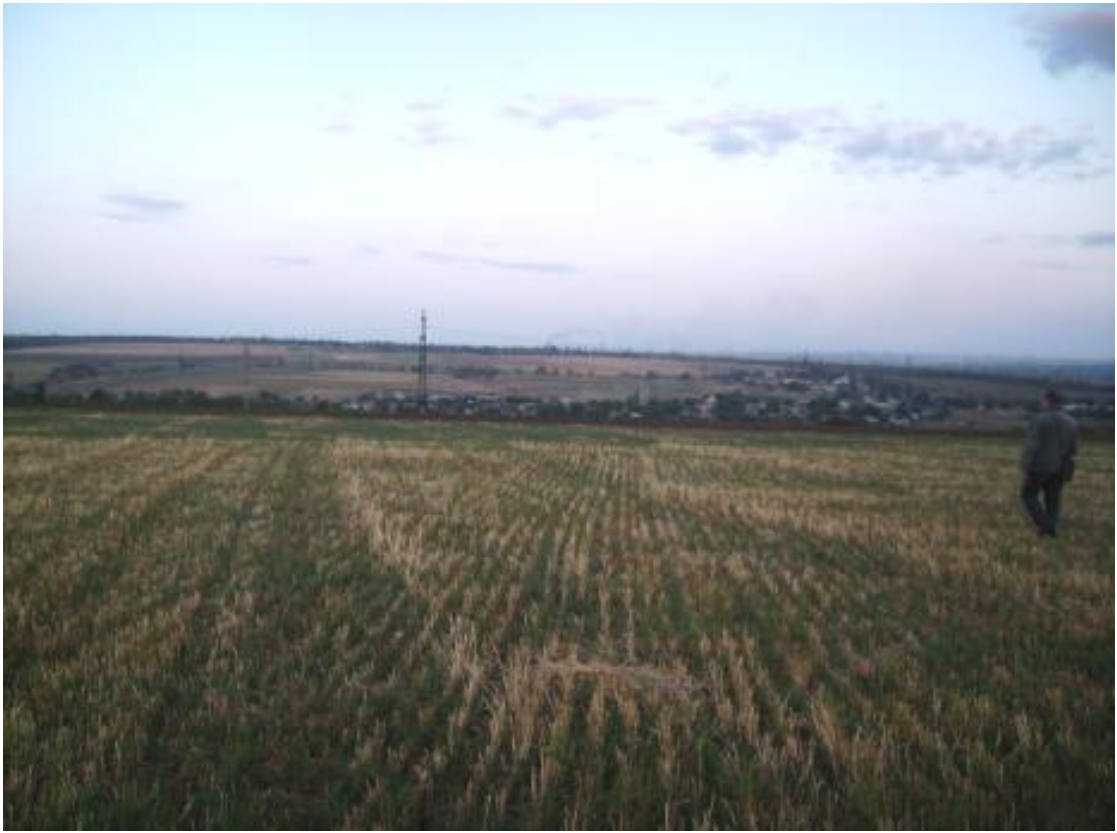

**Figure 9.** View of the Staropetrovsky village (foreground) in the vicinity from the ruined barrow, where the vessel with marks was found. On the horizon – pipes of Enakievo Steel (approx. 4.5 km), view from the west (photo A.Usachuka, September 2008). Photo made during archaeological research expedition of the Donetsk regional museum. In the photo on the right the member of the expedition – archaeologist V.A. Podobed.

Mark 9 shows conventional wells constructed as nail impression. Time marked that well corresponds to the height of Sirius +3º. The height of the Sun at the same time reaches about -6º. This position of the Sun characterizes the morning end of nautical twilight and the beginning of the civil twilight. From this point lighting is increased to such an extent that even bright navigational stars become invisible. Mark 9 shows the typical shape of corolla marks Staropetrovsky vessel. Likely to, it means that Sirius becomes at this moment already invisible as well. True local time of such relative position of Sirius and the Sun will change slightly for about ±500 years. In 900 BC such a situation arise, approximately, on 5 minutes later, and in 1900 BC – on 3 minutes earlier than 1400 BC (see. Table. 4).

As noted above, counting of the year from the date of heliacal rising of Sirius in Egypt was introduced at a time when Sirius ascend to a height of +3º at a time when the Sun reaches the height -6º. Mark 9 on the corolla fixes the just such a moment of time. Since the it is associated with the beginning of civil twilight and, accordingly, a qualitative change in the visible stars, we have tried to calculate the date and time when the achievement of by Sirius height +3º coincided with the beginning of astronomical and nautical twilight, because these moments are also associated with a qualitative change in the visibility of celestial objects. Beginning of astronomical twilight in the morning is characterized by the elevation of the Sun -18º. At this point the Milky Way is no longer visible. Remain visible only stars. After completion of astronomical twilight in the beginning of nautical twilight Sun reaches a height of -12º. From that moment the bulk of the stars in the sky become invisible. As mentioned above, only the most bright – navigational stars remain visible.

Our calculations show that the date for achieving by Sirius height +3º at the beginning of nautical twilight was 18 August. Moreover, the time of this astronomical situation coincided with the time corresponding to the edge of the mark 6, closer to the inner edge of the corolla. The date of achieving by Sirius height +3º at the beginning of astronomical twilight turned 26 August. Time of this astronomical situation coincided with the time corresponding to the edge of the mark 6, closer to the outer edge of the corolla. The calculations also show that on the same day – 26 August – Sirius ascended up, were at a height h≈0º, at a time corresponding to mark 5, which is the incised star. At this time, the Sun is at a height of about -20º, which corresponded to the period of astronomical night, which is characterized by the height of the Sun h <−18º. Lighting minimally at the astronomical night, so even the weakest objects in the sky become visible and accessible for visual observation, and Sirius began make an impression more bright stars, compared with the day of heliacal rising and the nearest days. Perhaps with this associated large size of star–shaped mark 5 compared with small star–shaped markers 7 and 8, the relevant date heliacal rising of Sirius.

Because a series marks associated with features of the apparent motion of Sirius for 17 days after the date of its heliacal rising, it is possible that the origins of the tradition of celebrating the holidays starting from the date of heliacal rising of Sirius [27] are connected with these days. Availability on the corolla of Staropetrovsky vessel a series of marks, associated with Sirius, allows for duplicate observations of Sirius near the day of its heliacal rising, for example, for correcting the date of the start of astronomical year or for the calculation of this date, if bad weather conditions prevented observations in the day of heliacal rising of Sirius, for example.

**Conclusion**

Thus, in the course of research for marks on the corolla of Staropetrovsky vessel, has developed a model of horizontal sundial with a sloping gnomon quite accurately describes the location of most of the corolla marks. The existence of this model testifies about possibilities use Staropetrovsky vessel not only as a water clock [4], but also as a sundial.

Several marks on the corolla of the vessel have star shape. Astronomical calculations show that their position on the corolla, as on "dial" of watch, indicates the time of qualitative change the visibility of Sirius in the day its heliacal rising and the next few days. The research results allow to state that astronomical year in carcass population began with the day heliacal rising of Sirius. Perhaps this tradition is the consequence of mediated Mesopotamian and/or Egyptian cultural influence.

## References

1. Klimenko, V.F.; Usachuk, A.N.; Tsymbal V.I. *Kurgannye drevnosti Central'nogo Donbassa. [Ancient mounds of the Central Donbass]*. Donetsk: RIP "Lebed", 1994, p. 102, 107–108.
2. Klimenko, V.F.; Tsymbal V.I. *O sosudax so znakami srubnoj obshhnosti e'poxi pozdnej bronzy v Severo–Vostochnom Priazov'e i Podoncov'e. [On vessels with signs carcass generality of the Late Bronze Age in the North East and the Sea of Azov Podontsove]*. DAS. Vol. 10. Donetsk: Donetsk National University Publishing House, 2002, p. 33.
3. Zakharova, E.Yu. *Sosudy so znakami srubnoj obshhnosti e'poxi pozdnej bronzy. [Vessels with signs carcass generality of the Late Bronze Age]*. Voronezh: TSCHKI, 2000, p. 89, 91.
4. Vodolazhskaya, L.N.; Usachuk, A.N.; Nevsky, M.Yu. Clepsydra of the Bronze Age from the Central Donbass. *Archaeoastronomy and Ancient Technologies* 2015, 3(1), 65–87.
5. Vodolazhskaya, L.N.; Vodolazhsky, D.I.; Ilyukov, L.S. Metodika komp'yuternoj fiksacii graficheskogo materiala arxeologicheskix raskopok na primere Karataevskoj kreposti. [Technique of of computer fixing graphic material of archaeological excavations on the example Karataevo fortress]. *Informacionnyj byulleten' Associacii Istoriya i komp'yuter. [Association Newsletter History and Computer]*, 2003, No 31, pp. 248–258
6. Vodolazhskaya L. Reconstruction of Heron's formulas for calculating the volume of vessels, in Posluschny, Axel; Lambers, Karsten; Herzog, Imela, Layers of Perception. *Proceedings of the 35th International Conference on Computer Applications and Quantitative Methods in Archaeology (CAA), Berlin, April 2–6, 2007*, Kolloquien zur Vor– und Frühgeschichte, Bd. 10, Berlin: Habelt; Propylaeum–DOC, 2008. – P. 1–7.
7. Vodolazhskaya, L.N.; Ponomarenko, V.O. Ximicheskij sostav keramiki svetloglinyanyx uzkogorlyx amfor varianta D tipa C IV iz Tanaisa. [The chemical composition of the ceramic svetloglinyanyh narrow–necked amphorae type D variant of the C IV Tanais]. *Trudy III (XIX) Vserossijskogo arxeologicheskogo s"ezda. [Proceedings of III (XIX) All–Russian Archaeological Congress]*. Veliky Novgorod – Staraya Russa. T. I St. Petersburg, Moscow, Novgorod the Great: 2011, 307–308.
8. Ponomarenko, V.O.; Sarychev, D.; Vodolazhskaya, L.N. Primenenie rentgenofluorescentnogo analiza dlya issledovaniya ximicheskogo sostava amfornoj keramiki. [The use of X–ray fluorescence analysis for the study of the chemical composition amphorae ceramic]. *Vestnik Yuzhnogo Nauchnogo Centra RAN. [Bulletin of the Southern Scientific Center, Russian Academy of Sciences]*. 2012, vol. 8, No 1, pp. 9–17
9. Vodolazhskaya, L.N.; Nevsky, M.Yu. Arxeoastronomicheskie issledovaniya svyatilishha Karataevo–Livencovskogo kompleksa. [Arhaeoastronomical research sanctuary

Karataevo–Liventsovsky complex]. *Metodika issledovaniya kul'tovyx kompleksov. [Technique of study of religious complexes].* Barnaul: OOO Pyat' plyus, 2012, pp. 5–13.

10. Vodolazhskaya, L.N.; Larenok, V.A. Arhaeoastronomical analysis of Levinsadovka sacrificial complex (South Russia). *Archaeoastronomy and Ancient Technologies* 2013, 1(1), 5–25.
11. Vodolazhskaya, L.N.; Nevsky, M.Yu. Russian meteorite of the Bronze Age (rock record). *Archaeoastronomy and Ancient Technologies* 2013, 1(2), 18–32.
12. Vodolazhskaya, L.N. Reconstruction of ancient Egyptian sundials. *Archaeoastronomy and Ancient Technologies* 2014, 2(2), 1–18.
13. Vodolazhskaya, L.N. Analemmatic and horizontal sundials of the Bronze Age (Northern Black Sea Coast). *Archaeoastronomy and Ancient Technologies* 2013, 1(1), 68–88.
14. Vodolazhskaya, L.N.; Larenok, P.A.; Nevsky, M.Yu. Ancient astronomical instrument from Srubna burial of kurgan field Tavriya–1 (Northern Black Sea Coast). *Archaeoastronomy and Ancient Technologies* 2014, 2(2), 31–53.
15. Zwicky, F. *Discovery, Invention, Research – Through the Morphological Approach.* Toronto: The Macmillan Company. 1969. 276 p.
16. Savoie, D. *Sundials design construction and use.* Springer, 2009. P. 71.
17. Waugh, A.E. *Sundials: Their Theory and Construction.* Dover Publications, Inc.: New York, USA, 1973. – PP. 35–51.
18. Kudryavtsev, V.A.; Demidovich, B.P. *Kratkij kurs vysshej matematiki. [Short course of higher mathematics].* M., 1978. P. 546.
19. Chubatenko, I.A.; Gorbov, V.N. Dannye o roste specializacii keramicheskogo proizvodstva srubnogo naseleniya Priazov'ya. [Data on the growth of specialty ceramic manufacturing carcass population of Azov region]. *Srubnaya kul'turno–istoricheskaya oblast'. Materialy III Rykovskix chtenij. [Srubna cultural and historical area].* Materials III Rykovsky readings. Saratov: Saratov State University Publishing House, 1994. P. 57, 59.
20. Kurtik, G.E. Izmerenie vremeni i kalendari v Drevnej Mesopotamii (shumerskij period). [The measurement of time and calendars in ancient Mesopotamia (Sumerian period)]. *Voprosy istorii estestvoznaniya i texniki. [Questions of history of science and technology].* № 4. 2013. P. 22–40.
21. Kononovich, E.V.; Moroz, V.I. *Obshhij kurs astronomii. [The general course of astronomy].* M .: Editorial, 2004, 544 p.
22. Kurtik, G.E. *Zvezdnoe nebo Drevnej Mesopotamii: shumero–akkadskie nazvaniya sozvezdij i drugix svetil. [Starry sky Ancient Mesopotamia: Sumerian–Akkadian names of the constellations and other heavenly bodies].* Petersburg: Aletheia, 2007. – P. 2–43.
23. Sachs, A.; Hunger, H. Astronomical Diaries and Related Texts from Babilonia. Vol. I: Diaries from 652 B.C. to 262 B.C. *Verlag der Österreichischen Akademie der Wissenschaften,* Vienna, 1988. 377 p.
24. Gautschy, R. Der Stern Sirius in Ägypten. *Zeitschrift für Ägyptische Sprache und Altertumskunde* 178, Vol. 2, 2011. P.116–131.
25. Parker, R. A. The Sothic Dating of the Twelfth and Eighteenth Dynasties. *Studies in Honor of G.R. Hughes (Studies in Ancient Oriental Civilization 39).* Chicago: The Oriental Institute, 1976. P. 182.

26. Berlev, O.D. Dva perioda Sotisa mezhdu Godom 18 carya Senu, ili Tosortrosa, i Godom 2 faraona Antonina Piya. [Two Sothis period between the year 18 of the king the Seine, or Tosortrosa and Year 2 Pharaoh Antoninus Pius]. *Drevnij Egipet: yazyk – kul'tura – soznanie. Po materialam egiptologicheskoj konferencii 12 – 13 marta 1998 g. [Ancient Egypt: language – culture – consciousness. According to the materials of the conference Egyptological 12 – 13 March 1998].* Moscow, 1999. – P. 42–62.
27. Holberg, J.B. *Sirius: Brightest Diamond in the Night Sky.* Chichester, UK: Praxis Publishing. 2007. 250 p.